\documentclass[12pt]{article}

\makeatletter
\renewcommand\section{\@startsection {section}{1}{\z@}%
                                   {-3.5ex \@plus -1ex \@minus -.2ex}
                                   {2.3ex \@plus.2ex}%
                                   {\normalfont\large\bfseries}}
\renewcommand\subsection{\@startsection{subsection}{2}{\z@}%
                                     {-3.25ex\@plus -1ex \@minus -.2ex}%
                                     {1.5ex \@plus .2ex}%
                                     {\normalfont\bfseries}}
\makeatother

\def\IZ{\relax\ifmmode\mathchoice
{\hbox{\cmss Z\kern-.4em Z}}{\hbox{\cmss Z\kern-.4em Z}}
{\lower.9pt\hbox{\cmsss Z\kern-.4em Z}} {\lower1.2pt\hbox{\cmsss
Z\kern-.4em Z}}\else{\cmss Z\kern-.4em Z}\fi}
\def\IR{\relax{\rm I\kern-.18em R}}

\def\one{{\hbox{ 1\kern-.8mm l}}}

\def\R{{\rm R}}

\def\tr{{\rm tr\,}}

\newlength{\bredde}
\def\slash#1{\settowidth{\bredde}{$#1$}\ifmmode\,\raisebox{.15ex}{/}
\hspace*{-\bredde} #1\else$\,\raisebox{.15ex}{/}\hspace*{-\bredde}
#1$\fi}

\newsavebox{\zzzbar}
\sbox{\zzzbar}
  {\setlength{\unitlength}{0.9em}
  \begin{picture}(0.6,0.7)
  \thinlines
  \put(0,0){\line(1,0){0.6}}
  \put(0,0.75){\line(1,0){0.575}}
  \multiput(0,0)(0.0125,0.025){30}{\rule{0.3pt}{0.3pt}}
  \multiput(0.2,0)(0.0125,0.025){30}{\rule{0.3pt}{0.3pt}}
  \put(0,0.75){\line(0,-1){0.15}}
  \put(0.015,0.75){\line(0,-1){0.1}}
  \put(0.03,0.75){\line(0,-1){0.075}}
  \put(0.045,0.75){\line(0,-1){0.05}}
  \put(0.05,0.75){\line(0,-1){0.025}}
  \put(0.6,0){\line(0,1){0.15}}
  \put(0.585,0){\line(0,1){0.1}}
  \put(0.57,0){\line(0,1){0.075}}
  \put(0.555,0){\line(0,1){0.05}}
  \put(0.55,0){\line(0,1){0.025}}
  \end{picture}}
\newcommand{\Tr}{\mathop{\mbox{Tr}}\nolimits}

\newcommand{\ena}{\end{eqnarray}}
\newcommand{\beqa}{\begin{eqnarray}}
\newcommand{\eeqa}{\end{eqnarray}}
\newcommand{\bea}{\begin{eqnarray}}
\newcommand{\eea}{\end{eqnarray}}

\newcommand{\eq}[1]{(\ref{#1})}

\newcommand{\be}{\begin{equation}}
\newcommand{\ee}{\end{equation}}

\usepackage{graphicx}

\newcommand{\ex}[1]{\mbox{e}^{\,\textstyle#1}}
\newcommand{\beq}{\begin{equation}}
\newcommand{\eeq}{\end{equation}}
\newcommand{\ber}{\begin{array}}
\newcommand{\eer}{\end{array}}

\newcommand{\del}{\partial}

\newcommand{\dsty}{\displaystyle}

\newcommand{\de}{\delta}

\newcommand{\cnst}{\mbox{const}}

\usepackage{amsmath}
\usepackage{amssymb}

\begin{document}
\begin{titlepage}
\begin{flushright}
\phantom{arXiv:yymm.nnnn}
\end{flushright}
\begin{center}
{\Large\bf Adiabaticity and emergence of classical space-time\vspace{2mm}\\
in time-dependent matrix theories}    \\
\vskip 10mm
{\large Ben Craps$^a$ and Oleg Evnin$^b$}
\vskip 7mm
{\em $^a$ Theoretische Natuurkunde, Vrije Universiteit Brussel and\\
The International Solvay Institutes\\ Pleinlaan 2, B-1050 Brussels, Belgium}
\vskip 3mm
{\em $^b$ Institute of Theoretical Physics, Academia Sinica\\
Zh\=onggu\=anc\=un d\=ongl\`u 55, Beijing 100190, China}
\vskip 3mm
{\small\noindent  {\tt Ben.Craps@vub.ac.be, eoe@itp.ac.cn}}
\end{center}
\vfill

\begin{center}
{\bf ABSTRACT}\vspace{3mm}
\end{center}

We discuss the low-curvature regime of time-dependent matrix
theories proposed to describe non-perturbative quantum gravity
in asymptotically plane-wave space-times. The emergence of near-classical space-time 
in this limit turns out to be closely linked
to the adiabaticity of the matrix theory evolution. Supersymmetry restoration at low curvatures,
which is crucial for the usual space-time interpretation of matrix theories,
becomes an obvious feature of the adiabatic regime.

\vfill

\end{titlepage}

\section{Introduction}

Time-dependent matrix theories \cite{Craps:2005wd} have been introduced as an analogue of the Banks-Fischler-Shenker-Susskind flat space-time matrix theory \cite{Banks:1996vh} 
and of matrix string theory \cite{Motl:1997th, Banks:1996my, Dijkgraaf:1997vv} in an attempt to describe
non-perturbative quantum gravity in time-dependent, possibly highly curved
(or even singular) space-times. The original set-up of \cite{Craps:2005wd} has been
later extended in various directions \cite{Li:2005sz,Li:2005ti,Das:2005vd,Chen:2005mga,Robbins:2005ua,Das:2006dr,Martinec:2006ak,Chen:2006rm,Ishino:2006nx,Bedford:2007dd,Blau:2008bp}; in particular, a systematic generalization of the analysis to more general singular homogeneous plane-wave space-time backgrounds has appeared in
\cite{Blau:2008bp}. In close parallel to the flat space matrix theory conjectures, one may
expect these models to give a complete quantum-gravitational theory of
asymptotically plane-wave space-times.  

The usual construction of (time-independent) matrix theories \cite{Banks:1996vh} essentially relies on the type IIA superstring/M-theory duality conjecture. Namely,
one compactifies a light-like dimension in the background 11-dimensional Minkowski space-time of M-theory, i.e., performs a discrete light-cone quantization (DLCQ) \cite{Susskind:1997cw}. 
To make the construction more precise \cite{Seiberg:1997ad} (see also \cite{Sen:1997we}), one takes this
compact dimension to be slighly space-like. A large boost relates this theory
to M-theory with a manifestly space-like compactification on a very small circle.
The latter system is related to weakly coupled type IIA string theory via the type IIA superstring/M-theory duality conjecture. Furthermore, for $N$ units of momentum
on the DLCQ circle and finite energies in the original reference frame used for DLCQ,
$N$ D0-branes have to be present in the type IIA theory, and the only surviving degrees
of freedom are the massless open string states associated with the D0-branes. This set-up yields the matrix theory action. An analogue of this argument can be devised for the case of IIA superstring theory (rather than M-theory), yielding a matrix string action \cite{Motl:1997th, Banks:1996my, Dijkgraaf:1997vv} (rather than matrix quantum mechanics).

Because plane waves enjoy a light-like isometry, the Minkowski space arguments
can be generalized to plane-wave backgrounds, resulting in time-dependent
matrix theories and matrix string theories. Even though, in the context of these models, novel physics is expected 
to emerge in the high-curvature regions
of the time-dependent backgrounds, it is also important to understand in what
precise manner the dynamics approaches classical space-time when the curvatures become
small. 

A related issue is supersymmetry restoration. Supersymmetry is essential in the flat space matrix theory to protect the free propagation
of gravitons, which in turn underlies the conventional space-time interpretation.
In time-dependent matrix theories, supersymmetry is broken completely (whichever
part is not broken by the plane wave backround, will be broken by the light-like
momentum on the DLCQ circle). It is therefore crucial to understand why this
does not prevent space-time from forming (at least in the low-curvature regime),
an issue that has been raised since the original formulation of this class of
models in \cite{Craps:2005wd}.

Heuristically, the question of low-curvature dynamics has been addressed already in \cite{Craps:2005wd}:
it has been shown that, in an appropriate parametrization, the time-dependent
matrix string theory reduces to a 2-dimensional supersymmetric gauge theory
on a Milne space-time with metric
\be\label{confmet}
ds^2 = e^{2\eta} \left( -d\eta^2 + dx^2 \right),\qquad x\sim x+2\pi.
\ee
The supersymmetry is only broken by the identification $x\sim x+2\pi$,
and, since the radius of the Milne circle becomes large at late times (i.e., in
the low curvature regime), one could expect that, for a wide range of processes,
the supersymmetry breaking will become invisible. In \cite{Craps:2006xq}, the effective
potential for the matrix string variables has been computed in the weak
coupling expansion of the gauge theory. Even though the computation is not technically
valid at late times, formal extrapolation of the resulting expressions suggested
that the effective potential decays at late times, a feature that could be
indicative of supersymmetry restoration. Another attempt to study late time (low background
curvature) dynamics of the time-dependent matrix theories has been recently undertaken
in \cite{O'Loughlin:2010uh}. 

Our present objective is to re-address these issues in a maximally clear and simple
fashion. For the 11-dimensional case of \cite{Li:2005sz}, the matrix action previously
presented in the literature shows steep time dependences at late times. This makes
the emergence of a near-classical space-time quite puzzling, as it superficially suggests
a strong explicit supersymmetry breaking, among other things. We shall show that, treated in appropriate variables, the relevant
time-dependent matrix theory approaches its flat space counterpart at late times
(low curvatures). The manner of convergence is somewhat subtle, but, since a bound
on deviations from the flat space theories will be given, the issues of space-time
interpretation and supersymmetry restoration are automatically resolved. One can understand this situation in a different way: the superficially steep
time dependences in the original matrix theory action actually enter an adiabatic\footnote{Adiabaticity has recently surfaced \cite{bdry} in the context of quantum gravity in time-dependent backgrounds, though the precise setting differs substantially from ours.} regime at late times, which again connects the
dynamics to that of a time-independent matrix theory. For the 11-dimensional case,
this is simply an equivalent and less straightforward way to view the geometry-inspired
variable redefinition that eliminates the time-dependences at late times.

For the 10-dimensional case of \cite{Craps:2005wd, Blau:2008bp}, however, there appears to be no canonical variable redefinition
that eliminates the time dependence in the Lagrangian at late times. (For instance, the 10-dimensional theory of \cite{Craps:2005wd} can be seen as the 11-dimensional matrix theory described in the previous passage, with an additional compactification. However, if
we attempt to perform the same variable redefinition as in the 11-dimensional case,
the compactification radius becomes time-dependent, making the theory awkward to study.) Still, one can establish
the onset of an adiabatic regime in the late-time dynamics of these theories, which
again connects them to their flat-space counterparts. (In justifying adiabaticity for this case, we shall rely on the standard flat-space matrix string
theory conjectures \cite{Dijkgraaf:1997vv}, which are essential to a meaningful interpretation
of the time-dependent matrix theories in any case, and hence already implicitly assumed.)

Note that in the present paper it is not our objective to trace the
complete evolution of
states from early to late times and show that a near-classical space-time
emerges from a generic initial state (which we do not expect to be the
case).
Explaining the emergence of a near-classical space-time from initial
conditions
is a major problem in cosmology, which we do not address here.
Rather, we shall show that the dynamics of the relevant matrix theories
approaches their flat-space counterparts, if studied at late times. This
implies
that, given a state with a late-time near-classical space-time
interpretation, matrix theory
can consistently describe its further evolution.
Establishing this property is already non-trivial, and it is an essential
pre-requisite for a more thorough treatment of cosmological scenarios in
our   framework.

We shall start our exposition by analyzing the 11-dimensional time-dependent
matrix theory, followed by the 10-dimensional matrix string theory. In these
treatments, we shall perform the algebraic manipulations explicitly, leading
up to the derivation of the necessary dynamical bounds. In the last section,
we shall explain how our formal manipulations are related to quantum adiabatic theory.

\section{Matrix quantum mechanics}

We shall start by briefly reviewing the 11-dimensional (quantum-mechanical)
matrix theories introduced in \cite{Li:2005sz} as simpler analogues of the 
matrix string theories of \cite{Craps:2005wd}. The relevant 11-dimensional
(M-theory) background has the form
\beq
ds^2=e^{2\alpha x^+}\left(-2dx^+dx^-+(dx^i)^2\right)+e^{2\beta x^+}(dx^{11})^2,
\eeq
or, in terms of the light-like geodesic affine parameter $\tau=e^{2\alpha x^+}/2\alpha$,
\beq
ds^2=-2d\tau dx^-+2\alpha\tau(dx^i)^2+(2\alpha\tau)^{\beta/\alpha}(dx^{11})^2.
\label{bckgr}
\eeq
This metric satisfies the 11-dimensional supergravity equations of motion
if the constants $\alpha$ and $\beta$ are related as $\beta=-2\alpha$,
or $\beta=4\alpha$. The fact that these relations need to be imposed
will not be relevant for our present considerations (it is essential,
however, for the general consistency of the corresponding matrix theories).

Since translations in $x^-$ form an isometry of the above background,
the usual DLCQ argument (proposed in \cite{Seiberg:1997ad} and adapted to the
time-dependent case in \cite{Craps:2005wd}) can be applied. The result \cite{Li:2005sz}
is a matrix theory that can be expected to describe non-perturbative
quantum gravity in space-times asymptotic to (\ref{bckgr}). The bosonic
and fermionic parts of the matrix theory action, respectively,
have the following form:
\beq
\ber{l}
\dsty S_B=\int d\tau\text{Tr}\left\{\frac{\alpha\tau}{R}(D_\tau X^i)^2+\frac{(2\alpha\tau)^{\beta/\alpha}}{2R}
(D_\tau X^{11})^2- \frac{R}{4}(2\alpha\tau)^2[X^i,X^j]^2\right.\vspace{2mm}\\
\dsty\hspace{2cm}\left.-\frac{R}{2}(2\alpha\tau)^{1+\beta/\alpha} [X^i,X^{11}]^2\right\},\vspace{3mm}\\
S_F=\int d\tau\left\{i\theta^TD_\tau\theta -R\sqrt{2\alpha\tau}\theta^T\gamma_i
[X^i, \theta ]-R(2\alpha\tau)^{\beta/2\alpha}\theta^T\gamma_{11} [X^{11}, \theta ]\right\}.
\eer
\label{mlact}
\eeq

The problems we intend to discuss can be understood already at the level
of the action (\ref{mlact}). The space-time backgrounds implicit in
(\ref{mlact}) are supposed to feature a light-like singularity at $\tau=0$
and become progressively more classical at large $\tau$. Yet, the explicit
time dependences in (\ref{mlact}) superficially become more steep, if anything,
at large $\tau$. This issue certainly deserves further clarification. Additionally,
there are no supersymmetries explicit in (\ref{mlact}). Since supersymmetries
are crucial for the free propagation of well-separated gravitons (and hence, a
robust geometrical interpretation) in the flat space matrix theory, one should
attempt to find an analogue of supersymmetry in (\ref{mlact}) that would enforce
a similar type of dynamics.

To address these important issues, we first note that the metric (\ref{bckgr})
describes a plane wave and, with the coordinate transformation
\beq
u=\tau,\quad z^i=\sqrt{2\alpha\tau}x^i,\quad z^{11}=(2\alpha\tau)^{\beta/2\alpha}x^{11},\quad v=x^-+\frac{\alpha(z^i)^2+\beta(z^{11})^2}{4\alpha\tau},
\label{coord}
\eeq
it can be brought to the Brinkmann form
\beq
ds^2=-2du\,dv-\frac{\alpha^2(z^i)^2-(\beta^2-2\alpha\beta)(z^{11})^2}{(2\alpha u)^2}du^2+(dz^i)^2+(dz^{11})^2.
\label{brinkmn}
\eeq
This parametrization forces the metric to manifestly approach Minkowski space-time
for large values of the light-cone time, which strongly suggests that the large
time dynamics of the corresponding matrix theory will likewise approach the flat
space matrix theory, if treated in appropriate variables. (As we shall see below, the convergence towards this limit is somewhat subtle, but the na\"\i ve
expectation will prove well-grounded.)

The matrix theory corresponding to (\ref{brinkmn}) can be constructed from
`scratch', as the metric enjoys the same $v$-translation isometry as (\ref{bckgr}).
More straightforwardly, one can apply the matrix analogue\footnote{Note that there is no matrix variable corresponding to $v$, since $v$ becomes the DLCQ circle in the
standard formulation of the matrix theory. The remaining transformations
are linear, so the issue of matrix multiplication ordering never arises.} of the transformations (\ref{coord}) directly to (\ref{mlact}) to obtain
\beq
\ber{l}
\dsty S_B=\int d\tau\text{Tr}\left\{\frac{1}{2R}\left[(D_\tau Z^i)^2+
(D_\tau Z^{11})^2\right]- \frac{R}{4}\left[[Z^i,Z^j]^2+2 [Z^i,Z^{11}]^2\right]\right.\vspace{2mm}\\
\dsty\hspace{2cm}\left.-\frac{\alpha^2(Z^i)^2-(\beta^2-2\alpha\beta)(Z^{11})^2}{(2\alpha \tau)^2}\right\},\vspace{3mm}\\
S_F=\int d\tau\left\{i\theta^TD_\tau\theta -R\theta^T\gamma_i
[Z^i, \theta ]-R\theta^T\gamma_{11} [Z^{11}, \theta ]\right\}.
\eer
\label{mlactbrnk}
\eeq

Note that the Brinkmann form of the matrix action has 
previously appeared \cite{Blau:2008bp} in the literature (for the 10-dimensional case).
However, the action in \cite{Blau:2008bp} corresponds to bringing the 10-dimensional
metric to the Brinkmann form, not the 11-dimensional metric (as we have done
presently). Both metrics are of the plane-wave form.

The action (\ref{mlactbrnk}) only differs from the flat space matrix theory
by a term decaying as $1/\tau^2$, thus one may expect that the late time
dynamics will be approximated by the flat space matrix theory and admit
the usual space-time interpretation. However, the decay is quite slow
and one might be worried about whether it is sufficient to ensure convergence.

To illustrate these worries, one may look at the straightforward example
of a harmonic oscillator whose frequency depends on time as $1/t^2$:
\beq
\ddot x + \frac{k}{t^2}x=0.
\label{ho}
\eeq
The two independent solutions to this equation can be given as $t^a$ and $t^{1-a}$,
where $a$ is a $k$-dependent number. These two solutions are obviously quite
different from a free particle trajectory, even though the equation of motion
approaches that of a free particle at late times. The reason for this discrepancy
is the slow rate of decay of the second term in (\ref{ho}).

However, in a physical setting, one is only able to perform {\it finite time} experiments.
That is, one has to specify the initial values $x(t_0)=x_0$, $\dot x(t_0)=v_0$
and examine the corresponding solution between $t_0$ and $t_0+T$. The solution is given by
\beq
x(t) = \frac{x_0(1-a)-v_0t_0}{1-2a}\left(\frac{t}{t_0}\right)^a + \frac{v_0t_0-x_0a}{1-2a}\left(\frac{t}{t_0}\right)^{1-a}.
\eeq
One can then see that $x(t_0+T)=x_0+v_0T+O(T/t_0)$, i.e., it is approximated
by a free motion arbitrarily well if the experiment starts sufficiently late.

It may be legitimately expected that the {\it finite time} behavior of the full time-dependent matrix theory given by (\ref{mlactbrnk}) will be approximated arbitrarily
well by the flat space matrix theory at late times, just as in the above harmonic
oscillator example. We shall now prove it by constructing an elementary bound
on dynamical deviations due to a small time-dependent term in the Schr\"odinger
equation.

We start with the following Schr\"odinger equation:
\beq
i\frac{d}{dt}\left|\Phi\right>=\left(H_0+f(t)H_1\right)\left|\Phi\right>,
\eeq
where $H_0$ and $H_1$ are time-independent, and rewrite it in the interaction picture (with respect to $H_0$):
\beq
\left|\Phi\right>=e^{-iH_0(t-t_0)}\left|\xi\right>,\qquad
i\frac{d}{dt}\left|\xi\right>=f(t)e^{iH_0(t-t_0)}H_1e^{-iH_0(t-t_0)}\left|\xi\right>.
\eeq
We then proceed to consider
\beq
\ber{l}
\dsty\frac{d}{dt}\Big|\left|\xi(t)\right>-\left|\xi(t_0)\right>\Big|^2=-\frac{d}{dt}\left(\left<\xi(t_0)\right.\left|\xi(t)\right>+c.c.\right)\vspace{3mm}\\
\dsty\hspace{2cm}=-if(t)\left(\left<\xi(t_0)\right|e^{iH_0(t-t_0)}H_1e^{-iH_0(t-t_0)}\left|\xi(t)\right>-c.c.\right).
\eer
\eeq
Integrating this expression between $t_0$ and $t_0+T$ and making use of
standard inequalities for absolute values and scalar products, we obtain:
\beq
\ber{l}
\dsty\Big|\left|\xi(t_0+T)\right>-\left|\xi(t_0)\right>\Big|^2=-i\int\limits_{t_0}^{t_0+T}dt f(t) \left(\left<e^{iH_0(t-t_0)}H_1e^{-iH_0(t-t_0)}\xi(t_0)\right.\left|\xi(t)\right>-c.c.\right)\vspace{3mm}\\
\dsty\hspace{2cm}\le 2\int\limits_{t_0}^{t_0+T}dt |f(t)| \sqrt{\left(\left|e^{iH_0(t-t_0)}H_1e^{-iH_0(t-t_0)}\xi(t_0)\right>\right)^2}\sqrt{\left(\left|\xi(t)\right>\right)^2}\vspace{3mm}\\
\dsty\hspace{2cm}\le 2\left(\mbox{max}_{[t_0,t_0+T]}|f(t)|\right)\int\limits_{0}^{T}dt \sqrt{\left<\xi(t_0)\right|e^{iH_0t}H_1^2e^{-iH_0t}\left|\xi(t_0)\right>}.
\eer
\label{bound}
\eeq
Now, assume that $f(t)$ approaches 0 at large times and consider a fixed $|\xi(t_0)\rangle\equiv |\xi_0\rangle$ (so we consider the evolution with fixed duration $T$ of the same initial state $|\xi_0\rangle$ starting at different initial times $t_0$). In this case, the first factor
in the last line becomes arbitrarily small for large $t_0$, whereas the second
factor does not depend on $t_0$. We then conclude that, for sufficiently late times,
the finite time evolution of the state vector will be approximated arbitrarily
well by $\left|\xi(t)\right>=\cnst$, i.e., by the evolution with $f(t)$ set identically
to 0.

It is then a simple corollary of the above bound that the time-dependent
matrix theory dynamics becomes approximated arbitrarily well at late times
by the flat space matrix theory, and, in particular, the supersymmetry is
asymptotically restored (with all the usual consequences, such as protection
of the flat directions of the commutator potential, and free graviton propagation).

A note may be in order: even though (\ref{bound}) does show that for any given
experiment (fixed $\left|\xi(t_0)\right>\equiv \left|\xi_0\right>$ and fixed $T$) deviations from
the flat space matrix theory become arbitrarily small for sufficiently large
$t_0$, this by no means implies that the convergence is uniform with respect
to $\left|\xi_0\right>$, which affects the value of the second factor
in the last line of (\ref{bound}). This is as it should be: for any $t_0$
there will be some experiments that will be able to discriminate between
the time-dependent and flat cases (with a given precision), yet, such experiments
will have to become more and more specialized (and less and less possible)
at late times.

\section{Matrix string theory}

The 10-dimensional ``Matrix Big Bang'' matrix string theories \cite{Craps:2005wd, Blau:2008bp} essentially differ from
the 11-dimensional case we have considered above by the compactness of one of the space-time dimensions. This feature precludes a straighforward variable
redefinition that would reduce the Lagrangian to a sum of time-independent terms
and terms decaying at large times (because the compactification radius would become time-dependent in the new variables). We shall therefore need to resort directly
to adiabaticity-inspired arguments to establish the emergence of near-classical space-time far
away from the Matrix Big Bang singularity.

The aim of the construction of matrix string theories \cite{Motl:1997th, Banks:1996my, Dijkgraaf:1997vv}
is to develop a non-perturbative description of quantum gravity in 10 dimensions (with the perturbative limit of this description given by the usual perturbative type IIA string theory).
For the original time-dependent Matrix Big Bang
matrix string theory of \cite{Craps:2005wd}, the 10-dimensional geometry is asymptotic
to the linear dilaton configuration:
\be
\begin{aligned}
ds_{st}^2 &= -2 dy^+ dy^- + (dy^i)^2,\\
\phi &= -Qy^+.
\end{aligned}
\label{mbb}
\ee

To construct the matrix string theory for the background (\ref{mbb}), one first lifts the background (\ref{mbb}) to 11 dimensions via the usual conjecture of type IIA/M-theory correspondence. The resulting 11-dimensional space-time is
\be\label{11met}
ds^2 = e^{ 2 Q y^+/3 } \left(-2 dy^+ dy^- + (dy^i)^2\right) + e^{-4 Q y^+/3} (dy)^2,
\ee
where $y$ is a coordinate along the M-theory circle. This is followed by the DLCQ compactification of the light-like $v$-coordinate, interpreted as the M-theory circle of an ``auxiliary'' type IIA string theory. A T-duality \cite{Taylor:1996ik} then relates the resulting theory of D0-branes on a compact dimension, i.e., a BFSS-like matrix theory with a compactified dimension, to a more manageable theory of wrapped D1-branes. This procedure has been carried out (in a slightly different but equivalent way) in \cite{Craps:2005wd} and has been reviewed in \cite{Craps:2006yb} and \cite{Blau:2008bp}. The resulting matrix string action is 
\be
\label{sym} S = \frac{1}{2\pi \ell_s^2}
\int {\tr}\left( \frac12(D_\mu X^i)^2 + \theta^T
{D\!\!\!\!\slash{}} \,\theta + \frac1{4g_{YM}^2} F_{\mu\nu}^2 - g_{YM}^2[X^i,X^j]^2 + g_{YM} \theta^T\gamma_i
[X^i,\theta]\right),
\ee
with the Yang-Mills coupling $g_{YM}$ related to the worldsheet values of the dilaton:
\be
g_{YM}=\frac{\ex{-\phi(y^+(\tau))}}{2\pi l_sg_s}=\frac{e^{Q\tau}}{2\pi l_sg_s}.
\ee

A generalization of this set-up has been proposed \cite{Blau:2008bp}. One can start with a 
10-dimensional power-law plane wave:
\beq
\begin{aligned}
\label{bn1}
ds_{st}^2 &= -2 dy^+ dy^- + g_{ij}(y^+) dy^i dy^j\equiv-2 dy^+ dy^- + \sum_{i} (y^+)^{2m_i} (dy^i)^2\\
&= -2dz^+dz^- + \sum_a \frac{m_a(m_a-1)}{(z^+)^2}(z^a)^2 (dz^+)^2 
+ \sum_a (dz^a)^2,\\
\ex{2\phi} &= (y^+)^{3b/(b+1)}=(z^+)^{3b/(b+1)}.
\end{aligned}
\eeq
Here, the first and the second line represent the Rosen and the Brinkmann form of the same plane wave, respectively. In order for the supergravity equations of motion to be satisfied, one needs to impose \cite{Blau:2008bp}
\be
\label{ede2}
\sum_i m_i(m_i-1) = -\frac{3b}{b+1}.
\ee
The original background of \cite{Craps:2005wd} can be seen as a $b\to-1$ limit of the above space-time \cite{Blau:2008bp}.

The 11-dimensional space-time corresponding to (\ref{bn1}) is
\be
\label{rcpl2}
\begin{aligned}
ds_{11}^2 &= -2dudv + \sum_i u^{2n_i}(dy^i)^2 + u^{2b}(dy)^2\\
&= - 2du dw + \sum_a \frac{n_a(n_a-1)}{u^2}(x^a)^2 (du)^2 +
\frac{b(b-1)}{u^2}x^2 (du)^2 + \sum_a (dx^a)^2 + (dx)^2, \\
\end{aligned}
\ee
with $n_i$ related to $m_i$ by $2m_i = (2n_i+b)/(b+1)$. Note that, since the Rosen and Brinkmann coordinates (first and second line in the above formula) are related by a $u$-dependent rescaling of the transverse coordinates, the identification of the (compact) $x$-variable is $u$-dependent. Thus, even though the second line of (\ref{rcpl2})
approaches a flat space-time at large values of $u$, the time-dependent identification
of $x$ makes an immediate application of the derivations of the previous section impossible.

The usual formulation leads to a matrix string action, whose bosonic part is given, in the Rosen coordinates of (\ref{bn1}), by
\begin{multline}
\label{src}
S_{RC} = \int d\tau d\sigma\Tr\left( -\frac{1}{4}g^{-2}_{YM}
\eta^{\alpha\gamma}\eta^{\beta\delta}F_{\alpha\beta}F_{\gamma\delta}
-\frac{1}{2}\eta^{\alpha\beta}g_{ij}(\tau)D_{\alpha}X^i D_{\beta}X^j
\right. \\ \left. 
+ \frac{1}{4}g^2_{YM} g_{ik}(\tau)g_{jl}(\tau)[X^i,X^j][X^k,X^l] \right)
.
\end{multline}
with the transverse metric $g_{ij}$ given by the first line of (\ref{bn1}) and the Yang-Mills coupling by
\be
g_{YM}=\frac{\ex{-\phi(y^+(\tau))}}{2\pi l_sg_s}=\frac{\tau^{-3b/2(b+1)}}{2\pi l_sg_s}.
\ee
One can further transform this action to the Brinkmann coordinates of the 
original plane wave, given by the second line of (\ref{bn1}), to obtain \cite{Blau:2008bp}:
\begin{multline}
\label{sbc}
S_{BC} = \int d\tau d\sigma\Tr\left( -\frac{1}{4}g^{-2}_{YM}
F_{\tau\sigma}^2
-\frac{1}{2}\left(D_{\tau}Z^a D_{\tau}Z^a-D_{\sigma}Z^a D_{\sigma}Z^a\right)\right. \\
\left. + \frac{1}{4}g^2_{YM} [Z^a,Z^b][Z^a,Z^b]  +
\frac{1}{2}A_{ab}(\tau)Z^aZ^b \right),
\end{multline}
where $A_{ab}=\mbox{diag}\{m_a(m_a-1)\}/\tau^2$. The latter form of the action only differs from a SYM gauge theory with a time-dependent coupling by the term involving
$A_{ab}$. The contribution of this term at late times can be shown to be negligible
by an argument very similar to the one employed for the 11-dimensional case in the previous section.

We thus end up at late times, both for the original Matrix Big Bang of \cite{Craps:2005wd} and for its generalization \cite{Blau:2008bp}, with a super-Yang-Mills theory with a growing time-dependent coupling that is a power-law or exponential function of time, and we have to analyze this
particular large time large coupling limit. Again, a puzzling feature here is that the time dependences
remain steep at large values of $\tau$ (set equal to $y^+$ by the gauge choice), 
far away from the singularity
of the original plane wave. The question of what happens at late times
has been discussed in the literature \cite{Craps:2005wd, O'Loughlin:2010uh} without definitive
quantitative answers provided.

The large coupling limit of the {\em time-independent} super-Yang-Mills theory is essential to the gravitational interpretation
of the flat space matrix string theory \cite{Dijkgraaf:1997vv}. Note that this limit is
distinct from the one we have to consider (time-dependent versus time-independent theory), however, in the adiabatic regime that our considerations will establish, properties of the time-dependent theory are related to properties of
time-independent ``snapshots'' of the time-dependent
Hamiltonian. We shall now assume the standard conjectures about
the large coupling behavior of the time-independent theory \cite{Dijkgraaf:1997vv}
and show that they result in an emergence of an adiabatic regime
within the time-dependent theory. The relevant manipulations will
be carried out in a pragmatic fashion, leading to a construction of the low-curvature limit.
We shall explain the relation of these manipulations to the general adiabatic
theory in the next section.

In \cite{Dijkgraaf:1997vv}, time-independent super-Yang-Mills theories have been conjectured to converge in the infrared ($g_{YM}\to\infty$) limit
to a sigma model conformal field theory on a permutation orbifold (with target space $S^N\R^8 \equiv (\R^8)^N/S_N$, where $S_N$ permutes the $N$ copies of $\R^8$) at finite $N$ (and to second quantized free strings
on Minkowski space in the large $N$ limit). The considerations given in support of
this claim are not rigorous, so we shall have to make natural assumptions below
about which particular form of convergence is implied.

Firstly, we assume that convergence occurs at the level of eigenstates: we postulate
that there exists a family of eigenvectors $\left|\Psi_n(g_{YM})\right>$ of $H_{YM}(g_{YM})$ such that
\begin{eqnarray}
&\dsty\left< \{A,\theta(x)\}|\Psi_n(g_{YM})\right>=\Psi_n^{\mbox{\tiny perm.orb.}}+\sum\limits_{k=1}^{\infty}\frac{\de\Psi_{nk}}{g_{YM}^k},\\
&\dsty H_{YM}\left|\Psi_n(g_{YM})\right>=\left(E_n^{\mbox{\tiny perm.orb.}}+\sum\limits_{k=1}^{\infty}\frac{\de E_{nk}}{g_{YM}^k}\right)\left|\Psi_n(g_{YM})\right>\equiv E_n(g_{YM})\left|\Psi_n(g_{YM})\right>.
\label{bsis}
\end{eqnarray}
Here $\left| \{A,\theta(x)\}\right>$ symbolically denotes eigenstates of the canonical coordinates of the super-Yang-Mills theory (the details depend on the quantization procedure).
Their appearance in the above expressions is necessary to compare wave-vectors of super-Yang-Mills theories at different values of the coupling (which are technically
defined in different Hilbert spaces). Further, $\Psi_n^{\mbox{\tiny perm.orb.}}$ and $E_n^{\mbox{\tiny perm.orb.}}$ denote wave-functionals of energy eigenstates of the permutation orbifold conformal field theory and their corresponding energies. (The above set-up assumes that the spectrum at finite $g_{YM}$ is a continuous deformation of the limiting spectrum at $g_{YM}\to\infty$. This may not always be so, for example, a discrete spectrum
of bound states at finite $g_{YM}$ may merge in the continuum at $g_{YM}\to\infty$. However, we are not aware of any such states in the context of perturbative string theory,
nor do we anticipate that they could have any dramatic effect on our considerations.
We shall therefore assume (\ref{bsis}) for our present purposes.) Note also that our summation over $n$ is a symbolic notation that implies integration over the continuous spectrum and summation over all discrete eigenvector labels, and none of our derivations use any assumptions about discreteness of the spectrum.

If the Hilbert space were finite-dimensional, the above conditions would guarantee
that the dynamical behavior of the theory reaches the free string limit suggested
in \cite{Dijkgraaf:1997vv} at large values of $g_{YM}$. However, for an infinite-dimensional space of states,
additional conditions enforcing not-too-poor convergence of highly excited eigenstates
need to be imposed. Indeed, in a physical setting, in order to declare that one theory
approaches another in a certain limit, one needs to ascertain that the evolutions
of the two theories become arbitrarily close to each other for any finite energy
normalizable initial state and for any finite duration of the experiment. Furthermore,
this convergence needs to be uniform for all normalizable initial states with
energies below some fixed value and all experiments lasting less than some given duration
(for, were it not so, there would always exist some finite energy finite duration
experiments discriminating between the two theories with a finite precision).
We embody these conditions in a statement that
\be
\sum_n  c_n e^{-i E_n(g_{YM})T} \left< \{A,\theta(x)\}|\Psi_n(g_{YM})\right> = \sum_n c_n e^{-i E_n^{\mbox{\tiny perm.orb.}}T} \Psi_n^{\mbox{\tiny perm.orb.}} + \sum\limits_{k=1}^{\infty} \frac{\psi_k}{g_{YM}^k},
\label{packet}
\ee
with $\sum |c_n|^2=1$ and $\psi_k$ being a set of normalizable wave-functionals. We further impose that the norms of $\psi_k$ are uniformly bounded for any $T$ less than
a chosen fixed value and for any initial state energy $\sum E_n^{\mbox{\tiny perm.orb.}}|c_n|^2$ less than a chosen fixed value.

We shall assume something slightly stronger than (\ref{packet}). Namely, we shall assume that the Yang-Mills coupling constant in the energies and eigenvectors on the left-hand side of (\ref{packet}) can be sent to infinity independently, still resulting in a power series expansion:
\be
\sum_n  c_n e^{-i E_n(g_{1})T} \left< \{A,\theta(x)\}|\Psi_n(g_{2})\right> = \sum_n c_n e^{-i E_n^{\mbox{\tiny perm.orb.}}T} \Psi_n^{\mbox{\tiny perm.orb.}} + \sum\limits_{\stackrel{k,l=0}{(k,l)\neq(0,0)}}^{\infty} \frac{\psi_{kl}}{g_{1}^kg_{2}^l}, 
\label{packet1}
\ee
with the same type of uniformity specifications we made under (\ref{packet}). We shall further assume that (\ref{packet1}) can be differentiated with respect to $T$ without
losing uniformity, which yields (with $T$ set to 0 and $g_2$ sent to infinity):
\be
\sum_n  c_n (E_n(g_{YM})-E_n^{\mbox{\tiny perm.orb.}}) \Psi_n^{\mbox{\tiny perm.orb.}} =  \sum\limits_{k=1}^{\infty} \frac{\phi_{k}}{g_{YM}^k}.
\label{energy}
\ee
As before, all these conditions are identical to (\ref{bsis}) for any finite-dimensional space of states, but in an infinite-dimensional space,
they constrain the convergence (in $g_{YM}$) at large $n$. (These conditions
are still considerably milder than demanding uniformity of convergence in $n$,
which would be quite unphysical.) There is a possibility that the derivations
below can be made without relying on anything beyond (\ref{packet}), but we have not
been able to devise such an argument.

Now we turn to the Schr\"odinger equation of the time-dependent super-Yang-Mills theory
\be
i\frac{d}{dt}\left|\Psi\right>=H_{YM}(g_{YM}(t))\left|\Psi\right>
\ee
and expand the state vector in the basis given by (\ref{bsis}):
\be
\left|\Psi\right>=\sum\limits_n c_n(t)\left|\Psi_n(g_{YM}(t))\right>.
\label{expnsn}
\ee
This yields
\be
i\frac{dc_n}{dt}+i\sum\limits_{m} c_{m}(t)\dot g_{YM}\left<\Psi_n(g_{YM})\right|\frac{d}{dg_{YM}}\left|\Psi_m(g_{YM})\right>=E_n(g_{YM})c_n(t).
\label{expnded}
\ee
We can rewrite this equation as
\be
i\frac{dc_n}{dt}=E_n^{\mbox{\tiny perm.orb.}}c_n + \sum_m H_{nm} c_m(t),
\label{Hmn}
\ee
where $H_{nm}$ are bounded by a negative power of $g_{YM}$: 
\be
H_{nm}=O(1/g^\gamma_{YM}),
\label{Hbnd}
\ee
with the precise value of the power depending on whether $g_{YM}$ is a power-law or exponential function of $t$. Note that the relevant coupling dependence coming from the
second term on the left-hand side of (\ref{expnded}) is $\dot g_{YM}/g_{YM}^2$, whereas the energy corrections on the right-hand side of (\ref{expnded}) are of order $1/g_{YM}$.

The fact that the matrix elements $H_{mn}$ are decreasing functions of time
suggests (though by no means in a conclusive way) that the evolution approaches
that of the permutation orbifold CFT at late times. We shall now analyze (\ref{Hmn}) in more detail to establish that this convergence does indeed take place.

Consider first a Schr\"odinger equation of the form
\be
i\frac{d}{dt}|\Phi\rangle=(H_0+H_1(t))|\Phi\rangle,
\ee
where $H_0$ is time-independent\footnote{It is an interesting mathematical question, to which the authors do not know the answer, what precise conditions should be imposed on $H_1(t)$ in order to make its contribution to the evolution small. There are many notions of convergence for Hilbert space operators, and saying that $H_1(t)$ becomes small in some limit is vacuous, unless the precise manner of convergence is specified.}. In parallel to the derivations of the previous section, we can rewrite it in the interaction picture (with respect to $H_0$):
\beq
\left|\Phi\right>=e^{-iH_0(t-t_0)}\left|\xi\right>,\qquad
i\frac{d}{dt}\left|\xi\right>= e^{iH_0(t-t_0)}H_1(t)e^{-iH_0(t-t_0)}\left|\xi\right>.
\eeq
We then proceed to consider
\beq
\ber{l}
\dsty\frac{d}{dt}\Big|\left|\xi(t)\right>-\left|\xi(t_0)\right>\Big|^2=-\frac{d}{dt}\left(\left<\xi(t_0)\right.\left|\xi(t)\right>+c.c.\right)\vspace{3mm}\\
\dsty\hspace{2cm}=-i\left(\left<\xi(t_0)\right|e^{iH_0(t-t_0)}H_1(t)e^{-iH_0(t-t_0)}\left|\xi(t)\right>-c.c.\right).
\eer
\eeq
Integrating this expression between $t_0$ and $t_0+T$ and making use of
standard inequalities for absolute values and scalar products, we obtain:
\beq
\ber{l}
\dsty\Big|\left|\xi(t_0+T)\right>-\left|\xi(t_0)\right>\Big|^2=-i\int\limits_{t_0}^{t_0+T}dt  \left(\left<e^{iH_0(t-t_0)}H_1(t)e^{-iH_0(t-t_0)}\xi(t_0)\right.\left|\xi(t)\right>-c.c.\right)\vspace{3mm}\\
\dsty\hspace{2cm}\le 2\int\limits_{t_0}^{t_0+T}dt \sqrt{\left\|e^{iH_0(t-t_0)}H_1(t)e^{-iH_0(t-t_0)}\left|\xi(t_0)\right>\right\|^2}\sqrt{\left\|\,\left|\xi(t)\right>\right\|^2}\vspace{3mm}\\
\dsty\hspace{2cm}= 2\int\limits_{0}^{T}dt \sqrt{\left\|H_1(t+t_0)e^{-iH_0t}\left|\xi(t_0)\right>\right\|^2}.
\eer
\label{bnd}
\ee 
For the Schr\"odinger equation (\ref{Hmn}), the state vector is given by the set of numbers $\{c_{n}\}$, and we take $H_0$ to be the first term on the right-hand-side, and $H_1(t)$ to be the second term (the detailed expression for $H_1$ can be read off from (\ref{expnded})). We furthermore denote $\left|\xi(t_0)\right>=\{c_{n}^{(0)}\}$, and observe that
\be
\ber{l}
\dsty H_1(t+t_0)e^{-iH_0t}\left|\xi(t_0)\right>\vspace{3mm}\\
\dsty\hspace{5mm}\equiv\sum_m \left( \dot g_{YM}\left<\Psi_n(g_{YM}(t+t_0))\right|\frac{d}{dg_{YM}}\left|\Psi_m(g_{YM}(t+t_0))\right>\exp\left[-i E_m^{\mbox{\tiny perm.orb.}}t\right]c_m^{(0)}\right)\vspace{3mm}\\
\dsty\hspace{2cm}+\left(E_n(g_{YM}(t+t_0))-E_n^{\mbox{\tiny perm.orb.}}\right)\exp\left[-i E_n^{\mbox{\tiny perm.orb.}}t\right]c_n^{(0)}.
\eer
\ee
The norm of the first term (i.e., the sum over $n$ of the square of its absolute value) is equal to the norm of the Hilbert space vector 
\be
\dot g_{YM} \sum_m c_m^{(0)}\exp\left[-i E_m^{\mbox{\tiny perm.orb.}}t\right] \frac{d}{dg_{YM}}\left|\Psi_m(g_{YM}(t+t_0))\right>
\label{vec}
\ee
(we use $\sum_n \left|\Psi_n(g_{YM}(t+t_0))\right>\left<\Psi_n(g_{YM}(t+t_0))\right|=1$) and it is bounded (uniformly with respect to $t$) by the $g_2$-derivative
of the condition (\ref{packet1}) with $g_1\to\infty$ to be $O(\dot g_{YM}/g_{YM}^2)$. The norm of the second
term is bounded (uniformly with respect to $t$) by (\ref{energy}) to be $O(1/g_{YM})$. Since the bounds are uniform with respect to $t$, they can be immediately integrated in (\ref{bnd}).

We then conclude that
\be
\Big|\left|\xi(t_0+T)\right>-\left|\xi(t_0)\right>\Big|^2=T\,O(1/g^\gamma_{YM}(t_0)),
\ee
i.e., deviations from the free string on the permutation orbifold become arbitrarily small at large $t_0$. Just as in the previous section, supersymmetry restoration at late
times becomes a simple corollary of the above bound.

\section{Adiabaticity}

In the preceding sections we have derived bounds controlling the convergence towards the late-time limit of time-dependent matrix theories.
A reader familiar with quantum adiabatic theory (see, e.g., \cite{messiah}) will
immediately recognize the structure of our manipulations.
Indeed, for both 11-dimensional and 10-dimensional cases, we have related
the time-dependent theory to properties of time-independent theories
(flat space matrix theory, super-Yang-Mills theories), which is 
characteristic of the adiabatic approximation. In this section,
we shall briefly review quantum adiabatic theory and display its
connections to the derivations in the previous sections.

Given a general quantum system with a time-dependent Hamiltonian $H(t)$
and a state vector $\left|\Psi\right>$ satisfying the Schr\"odinger equation
\beq
i\frac{d}{dt}\left|\Psi\right>=H(t)\left|\Psi\right>,
\eeq
one can always expand the state vector in a (time-dependent) basis
of {\it instantaneous} eigenvectors $\left|\Psi_n(t)\right>$ of $H(t)$:
\beq
\left|\Psi\right>=\sum\limits_n c_n(t)\left|\Psi_n(t)\right>,\qquad H(t)\left|\Psi_n(t)\right>=E_n(t)\left|\Psi_n(t)\right>.
\label{expansion}
\eeq
Note that our summation over $n$ is a symbolic notation that implies integration over the continuous spectrum and summation over all discrete eigenvector labels, and none of our derivations use any assumptions about discreteness of the spectrum. The Schr\"odinger equation then takes the form
\beq
i\frac{dc_n}{dt}+i\sum\limits_{m} c_{m}(t)\left<\Psi_n(t)\right|\frac{d}{dt}\left|\Psi_m(t)\right>=E_n(t)c_n(t).
\label{expanded}
\eeq

An adiabatic regime occurs when the second term on the left hand side becomes small.
Heuristically, this happens when $H(t)$ varies slowly in some sense, and so do
$\left|\Psi_n(t)\right>$, so that their derivatives can be neglected. In general,
it is difficult to spell out more handy conditions for the emergence of this
regime. However, in particular cases, simple and explicit adiabatic parameters
can be constructed, as we shall see below.

If the second term on the left hand side of (\ref{expanded}) can indeed be neglected,
the equations can be solved trivially to yield
\beq
c_n(t)=C_n \exp\left[-i\int dt E_n(t)\right],\qquad \left|\Psi\right>=\sum\limits_n C_n \exp\left[-i\int dt E_n(t)\right]\left|\Psi_n(t)\right>.
\label{genadisol}
\eeq
Note the close similarity between these approximate adiabatic solutions to
the time-dependent Schr\"odinger equation and the familiar solutions
to a time-independent Schr\"odinger equation: $\left|\Psi\right>=\sum_n C_n \exp\left[-iE_nt\right]\left|\Psi_n\right>$. (The stationary eigenvectors
$\left|\Psi_n\right>$ are simply replaced by the instantaneous
eigenvectors $\left|\Psi_n(t)\right>$, and the $E_nt$ in the phase factors
are simply replaced by $\int dt E_n(t)$.)

The relation between the time-dependent and time-independent systems becomes even more straightforward if the instantaneous spectrum $E_n(t)$
of $H(t)$ scales uniformly as a function of time, namely,
\beq
E_n(t)=\lambda(t)E^{(0)}_n,
\label{scaleE}
\eeq
where the $E^{(0)}_n$ do not depend on time. In that case, the phase factors
in (\ref{genadisol}) become simply $\left[-i E^{(0)}_n \int dt \lambda(t)\right]$,
in other words, they differ from the stationary case only by replacing $t$ with
$\int dt \lambda(t)$, {\it independently} of which state vector one is dealing with.

For the class of systems characterized by (\ref{scaleE}), it is convenient
to perform a variable redefinition in the Schr\"odinger equation. (\ref{scaleE})
implies that there exists a time-dependent unitary transformation $S(t)$
such that
\beq
S^\dagger(t)H(t)S(t)=\lambda(t)H_0,
\eeq
where $H_0$ does not depend on time and possesses the spectrum $E^{(0)}_n$.
One can then introduce $\left|\Phi\right>=S^\dagger(t)\left|\Psi\right>$, satisfying
\beq
i\frac{d}{dt}\left|\Phi\right>+i\left(S^\dagger(t)\frac{d}{dt}S(t)\right)\left|\Phi\right>=\lambda(t)H_0\left|\Phi\right>.
\label{schS}
\eeq
An adiabatic regime occurs when the second term on the left hand side can
be neglected compared to the right hand side. The time dependence
responsible for this relation between the two terms can be isolated
into the operator
\beq
\frac1{\lambda(t)}\left(S^\dagger(t)\frac{d}{dt}S(t)\right).
\eeq
If the second term on the left hand side of \eq{schS} can indeed be neglected,
the equation is solved to yield
\beq
\left|\Psi(t)\right>=S(t)\left|\Phi(t)\right>=S(t)\exp\left[-iH_0\int\limits_{t_0}^t dt \lambda(t)\right]S^\dagger(t_0)\left|\Psi(t_0)\right>
\label{Ssoln}
\eeq
The evolution is essentially that of a stationary system described by $H_0$,
except that a unitary transformation $S(t)$ is performed and the ``time flow''
is deformed from $t$ to $\int dt \lambda(t)$.

The situation simplifies further if $S(t)$ corresponds to a (time-dependent)
linear transformation of the canonical coordinates 
\be
q_k=\textbf{s}_{kl}(t)\tilde q_l
\label{linearad}
\ee
(and the corresponding transformation of the canonical momenta):
\beq
S(t)=\sqrt{\det\textbf{s}}\int dq \left|\textbf{s}\,q\right>\left<q\right|=\frac1{\sqrt{\det\textbf{s}}}\int dq \left|q\right>\left<\textbf{s}^{-1}q\right|.
\label{S}
\eeq
Differentiating $S(t)$ with respect to $t$ is somewhat subtle and is most conveniently
performed using
\be
\ber{l}
\dsty \frac{d}{dt}\left|\textbf{s}^{-1}q\right>\equiv\frac{d}{dt}\de(x-\textbf{s}^{-1}q)=
(-\textbf{s}^{-1}q)^{\phantom{.\hspace{-1mm}}^{\dsty\textbf{.}}}_k\,\del_k\de(x-\textbf{s}^{-1}q)\vspace{3mm}\\
\dsty\hspace{2cm}=i (\textbf{s}^{-1}\dot{\textbf{s}}\textbf{s}^{-1}q)_k\hat p_k\de(x-\textbf{s}^{-1}q)
\equiv i (\textbf{s}^{-1}\dot{\textbf{s}})_{kl}\hat p_k\hat q_l\left|\textbf{s}^{-1}q\right>.
\eer
\ee
Then,
\be
\frac{d}{dt}S=S\left(-\frac{(\det\textbf{s})^{\phantom{.\hspace{-1mm}}^{\dsty\textbf{.}}}}{2\det\textbf{s}}-i(\textbf{s}^{-1}\dot{\textbf{s}})_{kl}q_lp_k\right)=S\left(-\frac{i}2(\textbf{s}^{-1}\dot{\textbf{s}})_{kl}\{q_l,p_k\}\right).
\ee
Hence, (\ref{schS}) takes
the form
\beq
i\frac{d}{dt}\left|\Phi\right>+\frac12(\textbf{s}^{-1}\dot{\textbf{s}})_{kl}\{q_l,p_k\}\left|\Phi\right>=\lambda(t)H_0\left|\Phi\right>.
\label{lintranseq}
\eeq
All the time dependences have now been isolated into numerical pre-factors of the operators. The characteristic ratios (adiabatic parameters) quantifying the neglect 
of the second term on the left hand side with respect to the right hand side
can now be given\footnote{For the simple systems we are now considering, a purely classical consideration involving the transformation $q_k=\textbf{s}_{kl}(t)q'_l$ would produce the same adiabatic parameters. The way we have derived them here implies automatically their validity for the quantum case, which is
the physically relevant regime for gravitational matrix models.} as a matrix:
\beq
\frac{\textbf{s}^{-1}\dot{\textbf{s}}}{\lambda(t)}\quad.
\label{adparam}
\eeq

One can examine how the above formulas
work for the familiar case of a time-dependent harmonic oscillator with
$H(t)=(p^2+\omega^2(t)x^2)/2$. If one makes use of (\ref{S}) corresponding
to the transformation $x=\tilde x/\sqrt{\omega}$, one obtains $S^\dagger(t)H(t)S(t)=\omega(p^2+x^2)/2$. In other words, our above
analysis applies with $\textbf{s}=1/\sqrt{\omega}$ and $\lambda=\omega$,
so that (\ref{adparam}) becomes simply proportional to $\dot{\omega}/\omega^2$, i.e.,
the relative change of $\omega$ per period of the oscillations, which is
the familiar adiabatic parameter of the harmonic oscillator. The new Hamiltonian after the transformation has been effected can be read off (\ref{lintranseq}) as
\be
\tilde H = \frac{\omega}2 (p^2+x^2) +\frac{\dot\omega}{4\omega}(px+xp).
\label{hohaml}
\ee

We now turn to the derivations of the preceding sections. The variable redefinition (\ref{coord}) we have employed for the 11-dimensional matrix theory is precisely of the form (\ref{linearad}), and it converts the time-dependent system to a time-independent
system plus a correction decaying at large times. If we forget about the geometrical
interpretation of this variable redefinition, it simply becomes a particular case
of (\ref{linearad}) demonstrating that the (superficially steep) time dependences
in the Rosen form matrix theory action (\ref{mlact}) in fact become adiabatic at late
times, and the system is well approximated by its time-independent counterpart, i.e., the flat space matrix theory.

For the 10-dimensional case, equations (\ref{expnsn}-\ref{expnded}) are precisely of the form (\ref{expansion}-\ref{expanded}). Furthermore, (\ref{scaleE}) is almost satisfied at late times (because the spectrum approaches a constant limit). Our analysis of the 10-dimensional case shows explicitly that the second term on the left-hand side of (\ref{expnded}) can be neglected in comparison to the other terms (provided that the convergence of time-independent matrix string theories to the Dijkgraaf-Verlinde-Verlinde limit is sufficiently tame), which is by definition
the adiabatic regime.

One may try to object that adiabaticity is not a relevant term for our discussions,
since a variable redefinition brings the equations of motion to the form
where all the time dependent terms become small at large times.
This objection is vacuous, however, since it would also apply to a wide
range of systems commonly thought of as adiabatic. Indeed, if $E_n(t)$
in (\ref{expanded}) approaches a constant limit at late times, (\ref{expanded})
will take the form where all the time dependent terms become small at large times
whenever an adiabatic regime occurs at late times, in this general setting.
Likewise, the familiar time-dependent harmonic oscillator $H(t)=(p^2+\omega^2(t)x^2)/2$, the simplest
system used for text book demonstrations on adiabaticity, can be converted to
\be
H(t)=\frac{p^2+x^2}2+\frac{\dot\omega}{4\omega^2}(px+xp),
\label{hosc}
\ee 
if we start from (\ref{hohaml}) and introduce a new time variable $\tau= \int dt \,\omega(t)$ (the dot in the above formula still denotes the derivative with respect to the old time $t$ to maintain the familiar expression for the adiabatic parameter, $\dot\omega/\omega^2=\del_\tau\omega/\omega$). Then there is a one-to-one correspondence between the adiabatic
regime and the (new) Hamiltonian being almost constant: both occur when the adiabatic
parameter $\dot\omega/\omega^2$  is small. Yet, this mathematical structure does not
prevent anyone from employing the term `adiabatic' for the adiabatic regime of a time-dependent harmonic oscillator.

There appears to be widespread intuition that adiabaticity in quantum mechanics
is somehow connected to discreteness of the energy spectrum, and adiabatic
parameters emerge from comparisons of the rate of change of various terms in
the Hamiltonian to energy spacings in the discrete spectrum. This intuition
stems from the simplest versions of adiabatic theorems (proved, for example,
in \cite{messiah}), which are sufficient, but certainly not necessary conditions
for adiabaticity\footnote{One must also keep in mind that there are different ways
to reach the limit of slow relative variation of the Hamiltonian, and they may result
in different adiabatic parameters. For example, in adiabatic theorems proved in \cite{messiah}, one considers a {\it general} Hamiltonian $H(sT)$ where $s$ changes between
0 and 1, and the limit of slow variation is reached by sending $T$ to infinity.
The evolution is examined between $sT=0$ and $sT=T$, and demanding adiabaticity
is a very strong requirement, since the non-adiabatic terms should produce
a negligible contribution even over the huge time interval T. We study adiabatic
evolution on finite time intervals for Hamiltonians of a very particular form.}.
To dispel the doubts regarding adiabaticity in systems with a continuous
spectrum, one may simply notice that an inverted harmonic oscillator $H(t)=(p^2-\omega^2(t)x^2)/2$ can be brought to the form
\be
H(t)=\frac{p^2-x^2}2+\frac{\dot\omega}{4\omega^2}(px+xp)
\label{iho}
\ee 
by the same transformations we used in obtaining (\ref{hosc}). This system will
be adiabatic whenever $\dot\omega/\omega^2$ is small by virtue of the bound (\ref{bound}), irrespectively of the fact
that the spectrum is entirely continuous. For the matrix theories we have explored in this article, adiabaticity has been proved by constructing explicit bounds on deviations
from the strictly adiabatic evolution.

\section{Conclusions}

We have considered the low-curvature regime of time-dependent matrix theories and matrix string theories, and displayed the relation between the emergence of near-classical space-time and adiabaticity of the time dependences in the matrix theory actions.
In this context, supersymmetry of the matrix theories (explicitly broken by the
time dependence) is naturally restored at low curvatures, and the conventional
space-time interpretation of matrix theories becomes viable.

\section{Acknowledgments}
 
We would like to thank Matthias Blau, Sumit Das, Elias Kiritsis, Tim Nguyen, Martin O'Loughlin, Savdeep Sethi, Kostas Skenderis and Erik Verlinde for valuable discussions. This research has been
supported in part by the Belgian Federal Science Policy Office through the Interuniversity Attraction Pole IAP VI/11 and by FWO-Vlaanderen through project G011410N. The research of O.E.\ has also been supported by grants from the Chinese Academy of Sciences and National Natural Science Foundation of China. We would also like to thank the organizers of the Ascona {\em International Conference on Strings, M-Theory and Quantum Gravity} and of the {\it Focus week on strings and cosmology} (Institute for the Physics and Mathematics of the Universe, Tokyo) for hospitality while this work was nearing completion.


\end{document}